\documentstyle[aps,twocolumn]{revtex}
\title{Spectra of
Random Matrices Close to Unitary and Scattering Theory for
Discrete-Time Systems}
\author{ Yan V. Fyodorov}
\address{
Fachbereich Physik, Universit\"at-GH Essen,
D-45117 Essen, Germany}
\date{\today}
\begin{document}
\draft
\maketitle
\bigskip

\begin{abstract}

We analyze statistical properties of complex eigenvalues
of  random matrices $\hat{A}$ close to unitary. Such matrices appear
naturally when considering quantized chaotic maps within
a general theory of open linear stationary systems
with discrete time. Deviation from unitarity are characterized by
rank $M$ and eigenvalues $T_i, i=1,...,M$ of the matrix $\hat{T}=\hat{{\bf  
1}}-\hat{A}^{\dagger}\hat{A}$.
For the case $M=1$ we solve the problem completely by deriving
the joint probability density of eigenvalues and calculating all
$n-$ point correlation functions. For a general case we
present the correlation function of secular determinants.

\end{abstract}
\pacs{PACS numbers: 03.65 Nk, 05.45 Mt }
\vfill

The theory of wave scattering can be looked at as an integral part
of the general theory of linear dynamic open systems in terms of
the input-output approach. These ideas and relations
 were developed in system theory and
engeneering mathematics many years ago, see papers \cite{Liv,Ar,Hel}
and references therein. Unfortunately, that
 development went almost unnoticed by the majority of physicists
 working in the theory of chaotic quantum scattering and related
phenomena, see \cite{LW,FS} and references therein.
For this reason I feel it could be useful to
recall some basic facts of the input-output approach in
such a context.

An Open Linear System is characterized by three Hilbert spaces:
 the space $E_0$ of internal states $\Psi\in E_0$
and two spaces $E_{\pm}$ of incoming (-) and outgoing (+)
 signals or waves also called input and output spaces,
made of vectors $\phi_{\pm}\in E_{\pm}$. Acting in these three spaces
are four operators, or matrices: a) the so-called
fundamental operator $\hat{A}$ which maps any vector from internal
space $E_0$ onto some vector from the same space $E_0$ b)
two operators $\hat{W}_{1,2}$, with $\hat{W}_1$ mapping incoming
states onto an internal state and $\hat{W}_2$ mapping internal states onto
outgoing states and c) an operator $\hat{S}_0$ acting from $E_{-}$ to
$E_{+}$.

We will be interested in describing
the dynamics $\Psi(t)$ of an internal state with time $t$
provided we know the state at
initial instant $t=0$  and
the system is subject to a given input signal $\phi_{-}(t)$.
In what follows we consider only the case of
 the so-called stationary (or {\it time-invariant}) systems
when the operators are assumed to be time-independent.
Let us begin with the case of continous-time description.
The requirements of linearity, causality and
stationarity lead to a system of two dynamical equations:
\begin{equation}\label{1}
\begin{array}{c}
i\frac{d}{d t}\Psi=\hat{A} \Psi(t)+\hat{W}_1\phi_{-}(t)\\
\phi_{+}(t)=\hat{S}_0\phi_{-}(t)+i\hat{W}_2\Psi(t)
\end{array}
\end{equation}
Interpretation of these equations depends on the nature of state vector
$\Psi$ as well as of vectors $\phi_{\pm}$ and is different in different  
applications.
In the context of quantum mechanics one
relates the scalar product $\Psi^{\dagger}\Psi$ with the probability to find  
a particle inside the "inner" region at time $t$, whereas  
$\phi^{\dagger}_{\pm}\phi_{\pm}$ stays for probability currents flowing in  
and out of the region of internal states (the number of particles coming
or leaving the inner domain per unit time). The condition of particle  
conservation then reads as:
\begin{equation}\label{2}
\frac{d}{dt}\Psi^{\dagger}\Psi=
\phi^{\dagger}_{-}\phi_{-}-\phi^{\dagger}_{+}\phi_{+}
\end{equation}
It is easy to verify that Eq.(\ref{2})
 is compatible with the dynamics Eq.(\ref{1}) only provided the
operators satisfy the following relations:
\[
\hat{A}^{\dagger}-\hat{A}=i\hat{W}\hat{W}^{\dagger}\,\,,\,\,  
\hat{S}_0^{\dagger}\hat{S}_0=\hat{{\bf 1}}\ \, \mbox{and}\,
\hat{ W}^{\dagger}\equiv \hat{W}_2=-\hat{S}_0\hat{W}_1^{\dagger}
\]
which shows, in particular, that $\hat{A}$ can be written as  
$A=\hat{H}-\frac{i}{2}\hat{W}\hat{W}^{\dagger}$, with a Hermitian  
$\hat{H}=\hat{H}^{\dagger}$.

The meaning of $\hat{H}$ is transparent:
it governs the evolution $i\frac{d}{d t}\Psi=\hat{H} \Psi(t)$ of an
inner state $\Psi$ when the
coupling $\hat{W}$ between the inner space
and input/output spaces is absent. As such, it is
just the Hamiltonian describing
the closed inner region. The fundamental operator
$\hat{A}$ then has a natural
interpretation of the effective non-selfadjoint Hamiltionian describing
the decay of the probability from the inner region at
zero input signal: $\phi_{-}(t)=0$ for any $t\ge 0$.
If, however, the input signal is
given in the Fourier-domain by $\phi_{-}(\omega)$,
the output signal is related to it by:
\begin{equation}\label{3}
\phi_{+}(\omega)=\left[\hat{S}(\omega)\hat{S}_0\right]\phi_{-}(\omega)\quad,\quad  
\hat{S}(\omega)=
\hat{{\bf 1}}-i\hat{W}^{\dagger}\frac{1}{\omega \hat{{\bf 1}}-\hat{A}}\hat{W}
\end{equation}
where we assumed $\Psi(t=0)=0$. The unitary matrix $\hat{S}(\omega)$ is  
known in the mathematical
literature as the characteristic matrix-function of the non-Hermitian operator
$\hat{A}$. In the present context it is just the scattering matrix
whose unitarity is guaranteed by the conservation law Eq(\ref{2}).

The contact with the theory of chaotic scattering is
now apparent: the expression Eq.(\ref{3}) was
frequently used in the physical literature
\cite{LW,FS} as a starting point for extracting universal
properties of the scattering matrix for a quantum chaotic systems
 within the so-called random matrix approach.
The main idea underlying such an approach
is to replace the actual Hamiltonian
$\hat{H}$ by a large random matrix and to
calculate the ensuing statistics of the scattering matrix.
The physical arguments in favor of such a replacement can be found
in the cited literature.

In particular, most recently the
statistical properties of complex eigenvalues of
the operator $\hat{A}$ as well as related quantities
were studied in much detail\cite{FS,FK,SFT,Frahm}.
Those eigenvalues are poles of the scattering matrix and
have the physical interpretation as
{\it resonances} - long-lived intermediate states to which discrete
energy levels of the closed systems are transformed due
to coupling to continua.

In the theory presented above the time $t$ was a continous parameter.
On the other hand, a very useful instrument in the analysis of classical
Hamiltonian systems with chaotic dynamics are the so-called area-preserving
 chaotic maps\cite{Sm2}. They appear naturally either as a mapping
of the Poincare section onto itself, or as a result of stroboscopic
description of Hamiltonians which are periodic function of time.
Their quantum mechanical analogues are unitary operators which act on
Hilbert spaces of finite large dimension $N$. They are often referred
to as evolution, scattering or Floquet operators, depending on the
physical context where they are used. Their eigenvalues consist of
$N$ points on the unit circle (eigenphases). Numerical studies of
various classically chaotic systems suggest that the eigenphases
conform statistically quite accurately the results obtained for
unitary random matrices (Dyson circular ensembles).

Let us now imagine that a system represented by
 a chaotic map ("inner world") is embedded in a
larger physical system ("outer world") in such a way
that it describes particles which can
come inside the region of chaotic motion and leave
it after some time. Models of such type appeared, for example, in \cite{BGS}
where a kind of scattering theory for "open quantum maps" was developed
based on a variant of Lipmann-Schwinger equation.

On the other hand, in the general system theory \cite{Ar,Hel}
dynamical systems with discrete time are considered as frequently as
those with continous time.
 For linear systems  a "stroboscopic"  dynamics is just a linear map  
$(\phi_{-}(n); \Psi(n))\to(\phi_{+}(n); \Psi(n+1))$ which can be generally  
written as:
\begin{equation}\label{5}
\left(
\begin{array}{c}
\Psi(n+1)\\ \phi_{+}(n)
\end{array}
\right)
=\hat{V}
\left(\begin{array}{c}
 \Psi(n)\\ \phi_{-}(n)\end{array}\right)\quad,\quad
\hat{V}=\left(\begin{array}{cc}  
\hat{A}&\hat{W}_1\\\hat{W}_2&\hat{S}_0\end{array}\right)
\end{equation}
Again, we would like to consider a conservative system, and the
discrete-time analogue of Eq.(\ref{2}) is:
\[
\Psi^{\dagger}(n+1)\Psi(n+1)-\Psi^{\dagger}(n)\Psi(n)=
\phi^{\dagger}_{-}(n)\phi_{-}(n)-\phi^{\dagger}_{+}(n)\phi_{+}(n)
\]
which amounts to unitarity of the matrix $\hat{V}$ in Eq.(\ref{5}).
In view of such a unitarity $\hat{V}$
of the type entering Eq.(\ref{5}) can always be parametrized as (cf.\cite{Mel}):
\begin{equation}  \label{6}
\hat{V}= \left(\begin{array}{cc} \hat{u}_1&0\\0&\hat{v}_1\end{array}\right)
\left(\begin{array}{cc} \sqrt{1-\hat{\tau}\hat{\tau}^{\dagger}}& \hat{\tau}
\\\hat{\tau}^{\dagger}&-\sqrt(1-\hat{\tau}^{\dagger}\hat{\tau})
\end{array}\right)\left(\begin{array}{cc}
\hat{u}_2&0\\0&\hat{v}_2\end{array}\right)
\end{equation}
where the matrices $u_{1,2}$ and $v_{1,2}$ are unitary and
$\hat{\tau}$ is a rectangular $N\times M$ diagonal matrix with the entries
$\tau_{ij}=\delta_{ij}\tau_i, \, 1\le i\le N\,,\,  1\le j\le M\quad 0\le;  
\,\tau_i\le 1$.

Actually, it is frequently convenient to redefine
input, output and internal state
as: $\phi_{-}(n)\to \hat{v}_2^{-1}\phi_{-}(n),\quad \phi_{+}\to \hat{v}_1
\phi_{+}(n)$ and $\Psi(n)\to\hat{u}_2\Psi(n)$ which amounts just to
 choosing an appropriate basis in the corresponding spaces.
The transformations bring the matrix $\hat{V}$ to somewhat simplier
form:
\begin{equation}  \label{7}
\hat{V}=
\left(\begin{array}{cc} \hat{u} \sqrt{1-\hat{\tau}\hat{\tau}^{\dagger}}& u  
\hat{\tau}
\\ \hat{\tau}^{\dagger}&-\sqrt{1-\hat{\tau}^{\dagger}\hat{\tau}}\end{array}
\right)
\end{equation}
where $ \hat{u}= \hat{u}_2^{\dagger}\hat{u}_1$. Such a form suggests
clear interpretation of the constituents of the model. Indeed,
for $\hat{\tau}=0$ the dynamics of the system amounts to:
$\Psi(n+1)=\hat{u}\Psi(n)$. We therefore identify $\hat{u}$ as a unitary
evolution operator of the "closed" inner state domain decoupled both
from input and output spaces. Correspondingly,
$\hat{\tau}\ne 0$ just provides a coupling that makes the system
open and converts the fundamental operator $\hat{A}=\hat{u}
\sqrt{1-\hat{\tau}\hat{\tau}^{\dagger}}$ to a contraction:
$1-\hat{A}^{\dagger}\hat{A}=\tau\tau^{\dagger}\ge 0$. As a result,
the equation $\Psi(n+1)=\hat{A}\Psi(n)$ describes an irreversible
decay of any initial state $\Psi(0)$ for zero input $\phi_{-}(n)=0$,
whereas for a nonzero input and $\Psi(0)=0$ the Fourier-transforms
$\phi_{\pm}(\omega)=\sum_{n=0}^{\infty}e^{in\omega}\phi_{\pm}(n)$
 are related by a unitary scattering matrix $\hat{S}(\omega)$
given by:
\begin{equation} \label{8}
\hat{S}(\omega)=-\sqrt{1-\hat{\tau}^{\dagger}\hat{\tau}}
+\hat{\tau}^{\dagger}\frac{1}{e^{-i\omega}-\hat{A}}\hat{u}\hat{\tau}
\end{equation}

 Assuming further that the motion outside the inner region is regular, we  
should be able to describe generic features of open quantized chaotic maps
 choosing the matrix $\hat{u}$ to be a member
of a Dyson circular ensemble.
Then, averaging $\hat{S}(\omega)$ in Eq.(\ref{8}) over $\hat{u}$ one finds:  
$\hat{\tau}^{\dagger}\hat{\tau}=1-\left|\overline{\hat{S}
(\omega)}\right|^2$. Comparing this result with \cite{LW,FS} we see that
$M$ eigenvalues $0\le T_i\le 1$ of the $M\times M$ matrix  
$\hat{T}=\hat{\tau}^{\dagger}\hat{\tau}$ play a role of the so-called  
transmission coefficients and describe a
particular way the chaotic region is coupled to the outer world.

In fact, this line of reasoning is motivated by recent papers
\cite{Kol,KS}. The authors of \cite{Kol} considered the Floquet
description of a Bloch particle in a constant force and periodic
driving. After some approximations the evolution of the system is
described by a mapping: ${\bf c}_{n+1}={\bf F}{\bf c}_{n}$, where
the unitary Floquet operator ${\bf F}=\hat{S}\hat{U}$ is the product of a
unitary "M-shift"
$\hat{S}: S_{kl}=\delta_{l,k-M}, \, l,k=-\infty,\infty$
and a unitary matrix $\hat{U}$. The latter is
effectively of the form
$\hat{U}=\mbox{diag}(\hat{d_1},\hat{u},d_2)$, where $\hat{d}_{1,2}$
are (semi)infinite diagonal matrices
 and $\hat{u}$ can be
taken from the ensemble of random $N\times N$ unitary matrices.

One can check that such a dynamics can be easily brought to the standard
Eqs.(\ref{5},\ref{6}) with the fundamental operator being $N\times N$
random matrix of the form
$\hat{A}=\sqrt{{\bf 1}-\hat{\tau}^{\dagger}\hat{\tau}} \hat{u}$, and all
$M$ diagonal elements of $N\times M$ matrix $\tau$ are equal to
unity. Actually, the original paper \cite{Kol} employed a slightly
different but equivalent construction dealing with
an "enlarged" internal space of the dimension $N+M$. We prefer to
follow the general scheme because of its conceptual clarity.

Direct inspection immediately shows that the non-vanishing
 eigenvalues of the fundamental operator $\hat{A}$ as above coincide
with those of $(N-M)\times (N-M)$ subblock of the  random
unitary matrix $u$. Complex eigenvalues of such "truncations" of random
unitary matrices were studied in much detail by the authors of
a recent insightful paper \cite{KS}. They managed to study eigenvalue
correlations analytically for arbitrary $N,M$. In particular,
they found that in the limit $N\to \infty$ for fixed $M$ these
correlation functions practically coincide \cite{Kol} with those obtained
earlier \cite{FS,FK} for operators of the form $\hat{A}=\hat{H}-
\frac{i}{2}\hat{W}\hat{W}^{\dagger}$ occuring in the theory of
open systems with continous-time dynamics.

Such a remarkable universality, though not completely unexpected,
deserves to be studied in more detail. In fact, truncated unitary
matrices represent only a particular case of random
contractions $A$.  Actually, some statistical properties
of general subunitary matrices  were under investigation recently
as a model of scattering matrix for systems with absorption,
see \cite{abs}.

The main goal of the present paper
is to add to our knowledge on specta of random contractions
for a given deviation from unitarity.

The ensemble of general $N\times N$ random contractions
$\hat{A}=\hat{u}
\sqrt{1-\hat{\tau}\hat{\tau}^{\dagger}}$
can be described by the following probability measure in the matrix space:
\begin{equation}\label{0}
{\cal P}(\hat{A})
d\hat{A}\propto \delta(\hat{A}^{\dagger}\hat{A}-\hat{G}) dA\quad, \quad
\hat{G}\equiv 1-\hat{\tau}\hat{\tau}^{\dagger}
\end{equation}
where $dA=\prod_{ij} dA_{ij}dA^{*}_{ij}$.
The $N\times N$ matrix $\hat{\tau}\hat{\tau}^{\dagger}={\bf 1}- \hat{G}\ge  
0$ is natural to call the deviation matrix. It has $M$ nonzero eigenvalues  
coinciding with the transmission coefficients $T_a$ introduced above.
The particular choice $T_{i\le M}=1,\,\, T_{i> M}=0$
corresponds to the case considered in \cite{KS}. In what follows we assume  
all $T_i\le 1$.

Our first step is, following \cite{FK,KS}, introduce the Schur decomposition
$\hat{A}=\hat{U}(\hat{Z}+\hat{R})\hat{U}^{\dagger}$ of the matrix $A$ in  
terms of a unitary $\hat{U}$, diagonal matrix of the eigenvalues $\hat{Z}$  
and a lower triangular $\hat{R}$. One can satisfy oneself, that
the eigenvalues $z_1,...,z_N$ are generically not degenerate, provided
all $T_i<1$. Then, the measure written in terms of new variables is given by  
$d\hat{A}=|\Delta(\{z\})|^2 d\hat{R}d\hat{Z}d\mu(U)$, where
the first factor is just the Vandermonde determinant of eigenvalues $z_i$
and $d\mu(U)$ is the invariant measure on the unitary group.
The joint probability density of complex eigenvalues is then given by:
\begin{eqnarray}\label{9}
{\cal P}(\{z\})&\propto& |\Delta(\{z\})|^2 \\ \nonumber
\times & \int& d\mu(U)d\hat{R}\
\delta\left((\hat{Z}+\hat{R})(\hat{Z}+\hat{R})^{\dagger}-
\hat{U}^{\dagger}\hat{G}\hat{U}\right)
\end{eqnarray}
The integration over $\hat{R}$ can be performed with some manipulations
using its triangularity (some useful hints can be found in \cite{KS}).
As the result, we arrive at:
\begin{eqnarray}\label{10}
{\cal P}(\{z\})&\propto& |\Delta(\{z\})|^2 \\
\nonumber \times
\int d\mu(U) & \prod_{l=1}^N& \left.
\delta\left(|z_1|^2...|z_l|^2- \det\left[\left(
\hat{U}^{\dagger}\hat{G}\hat{U}\right)_{ij}\right]\right|_{(i,j)=1,...,l}\right)
\end{eqnarray}
The remaining integration over the unitary group poses a serious problem.
We found no way to overcome the difficulties in a general case. However,  
when the rank $M$ of the deviation matrix $\hat{\tau}\hat{\tau}^{\dagger}$ is  
unity, we managed to perform the integral by methods of \cite{FK} and  
arrived at a very simple expression:
\begin{equation}\label{11}
{\cal P}(\{z\})\propto T^{1-N}|\Delta(\{z\})|^2
\delta\left(1-\hat{T}-|z_1|^2...|z_N|^2\right)
\end{equation}
provided $0\le |z_l|\le 1$ for all eigenvalues, and zero otherwise.
Here $0<T<1$ is the only non-zero eigenvalue of the deviation matrix.
The Eq.(\ref{11}) can be used to extract all n-point correlation functions:
\begin{eqnarray}
R_n( z_1,...,z_n)=\frac{N!}{(N-n)!}\int d^2z_{n+1}...d^2z_N
{\cal P}(\{z\})
\end{eqnarray}
To achieve this it is convenient to use the Mellin transform with respect to  
the variable $\zeta = 1-T$:
\begin{equation}\label{12}
\tilde{R}_n(s;\{z\}_n)=\int_{0}^{\infty}d\zeta \zeta^{s-1}
\left[(1-\zeta)^{N-1}R_n(\{z\}_n)\right]
\end{equation}
It easy to notice that such a transform brings ${\cal P}(\{z\})$ to the form
suitable for exploitation of the orthogonal polynomial method
\cite{Mehta}. The corresponding polynomials are $p_k(z)=\sqrt(k+s)z^n$  
orthonormal with respect to the weight $f(z)=|z|^{2(s-1)}$ inside the unit  
circle $|z|\le 1$.
Following the standard route we find:
\begin{equation}\label{13}
\tilde{R}_n(s;\{z\}_n)\propto
\frac{\det{\left[K(z_i,z_j)\right]}|_{(i,j)=1,...,n}}{s(s+1)...(s+N-1)}
\end{equation}
where the kernel is
\begin{eqnarray}\label{13a}
K(z_1,z_2)&=&(f(z_1)f(z_2))^{1/2}\sum_{k=0}^{N-1}p_k(z_1^*)p_k(z_2)\\
\nonumber & = &
|x|^{s-1}\left( s\phi(x)+x\frac{d}{d x}\phi(x)\right)|_{x=z_1^*z_2}
\end{eqnarray}
and $\phi(x)=(x^N-1)/(x-1)$.
Thus, the expression  Eq.(\ref{13}) can be rewritten as:
\begin{eqnarray}\nonumber
\tilde{R}_n(s;\{z\}_n)&\propto& \frac{\prod_{l=1}^n|z_l|^2}{s(s+1)...(s+N-1)}
\sum_{l=0}^ns^l q_l(\{z\}_n);\\
\nonumber
q_0(\{z\}_n)&=&\det{\left[x\frac{d}{d  
x}\phi(x)|_{x=(z_i^*z_j)}\right]}|_{i,j=1,...,n}\\ \nonumber &\ldots &\\
\nonumber q_n(\{z\}_n)&=&\det{\left[\phi(x)|_{x=(z_i^*z_j)}\right]}|_{i,j=1,...,n}
\end{eqnarray}
and can be easily Mellin-inverted yielding finally the original correlation  
functions in the following form:
\begin{eqnarray}
R_n(\{z\}_n )|_{|z_n|\le 1}&\propto&
T^{1-N}\theta(T-1+a)\sum_{l=0}^n q_l(\{z\}_n) \\ \nonumber
\times \left(\frac{d}{da}a\right)^l && \left.
\left[\frac{1}{a}\left(1-\frac{1-T}{a}\right)\right]^{N-1}
\right|_{a=\prod_{i=1}^n|z_i|^2}
\end{eqnarray}
where $\theta(x)=1$ for $x\ge 0$ and zero otherwise.
This equation is exact for arbitrary $N$. Let us now investigate the limit
$N\gg n$ and use:
\[\left(\frac{d}{da}a\right)^l
\left[\frac{1}{a}\left(1-\frac{\xi}{a}\right)\right]^{N-1}\approx
\left(\frac{N\xi}{a-\xi}\right)^l\frac{1}{a}\left(1-\frac{\xi}{a}\right)^{N-1}
\]
which allows one to rewrite the correlation function as:
\begin{eqnarray}\label{20}
R_n(\{z\}_n )&\propto&
T^{1-N}\frac{1}{a}\left(1-\frac{1-T}{a}\right)^{N-1}
\theta(T-1+a) \\ \nonumber
&\times & \det_{i,j=1,...,n}{\left(\frac{N(1-T)}{a-1+T}\phi(x)
+x\frac{d}{d x}\phi(x)\right)|_{x=z_i^*z_j}}
\end{eqnarray}

Further simplifications occur after taking into account that
eigenvalues $z_i$ are expected to concentrate
typically at distances  of order of $1/N$ from the unit circle.
Then it is natural to introduce new variables
$y_i,\phi_i$ according to $z_i=(1-y_i/N) e^{i\phi_i}$
and consider $y_i$ to be of the order of unity when $N\to \infty$.
As to the phases $\phi_i$, we expect their typical separation scaling as:
$\phi_i-\phi_j=O(1/N)$.
 Now it is straightforward to perform the limit
$N\to \infty$ explicitly and bring Eq.(\ref{20}) to the final form:
\begin{eqnarray}\nonumber
R_n(\{z\}_n )&\propto&e^{-g\sum_{i=1}^n y_i}\det{\left[\int_{-1}^{1}d\lambda  
(\lambda+g)
e^{-\frac{i}{2}\lambda\delta_{ij}}\right]_{i,j=1,n}}
\end{eqnarray}
with $g=2/T-1$ and $\delta_{ij}=N(\phi_i-\phi_j)-i(y_i+y_j)$.
The expression above coincides in every detail with that obtained
in\cite{FK} for random matrices with rank-one deviation
from Hermiticity provided one remembers that mean linear
density of phases $\phi_i$ along the unit circle is $\nu=1/(2\pi)$.
This completes the proof of universality for rank-one deviations.

Being so far unable to evaluate the spectral correlation function for
an aritrary $\hat{A}$, we nevertheless succeeded in
calculating closely related but simpler
object, namely, the correlation function of secular determinants:
\begin{equation}\label{21}
I(z_1,z_2)=\left\langle(\det{\left(z_1{\bf 1}-\hat{A}\right)}
\det{\left(z_2^*{\bf 1}-\hat{A}^{\dagger}\right)}\right\rangle_A
\end{equation}
where the angular brackets stand for the averaging over the
probability density in Eq.(\ref{0}). Such a correlation function
for unitary $\hat{A}$ corresponding to quantum chaotic maps
attracted much attention recently \cite{Sm2,sec}. For
a non-Hermitian $\hat{A}$ similar objects were studied in\cite{FK}.

Relegating details to a more extended publication, we just present the
final result in terms of eigenvalues $g_i=1-T_i$ of the matrix $\hat{G}$:
\begin{eqnarray}\label{22}
I(z_1,z_2)&=&\\ \nonumber &&
\sum_{k=0}^N (z_1z_2^*)^k\left(\begin{array}{c}N\\k
\end{array}\right)^{-1}
\sum_{1\le i_1<i_2<...<i_k\le N} g_{i_1}g_{i_2}...g_{i_k}
\end{eqnarray}
For random unitary matrices all $T_i=0$ and the expression above reduces
to $I(z_1,z_2)=\sum_{k=0}^N(z_1z_2^*)^k$ in agreement with \cite{sec}.

The author is obliged to B. Khoruzhenko and especially
to H.-J. Sommers for stimulating discussions
and suggestions.
The financial support by SFB 237 "Unordnung und grosse Fluktuationen"
as well as of the grant No. INTAS 97-1342 is acknowledged with thanks.


\begin{references}
\bibitem[*]{leave} Address since March 2000:\\
Department of Mathematical Sciences, Brunel University, Cleveland Road,  
Uxbridge, UB8 3PH, UK
\bibitem{Liv} M.S.Livsic "Operators, Oscillations, Waves: Open
Systems." (Translations of Mathematical Monographs, v.34 (AMS,
Providence, RI, 1973)
\bibitem{Ar} D.Z.Arov {\it Sib.Math.Journ.} {\bf 20} (1979), 149
\bibitem{Hel} J.W.Helton {\it J.Funct.Anal.} {\bf 16} (1974), 15
\bibitem{LW} C.H.Lewenkopf and H.A.Weidenm\"{u}ller {\it Ann.Phys. NY}
{\bf 212} (1991), 53
\bibitem{FS} Y.V.Fyodorov and H.-J.Sommers {\it J.Math.Phys.} {\bf
38} (1997), 1918
\bibitem{FK} Y.V.Fyodorov and B.A.Khoruzhenko {\it Phys.Rev.Lett.}
{\bf 83} (1999), 65
\bibitem{SFT} H.-J.Sommers, Y.V.Fyodorov and M.Titov {\it J.Phys.A} {\bf 32}
(1999), L77
\bibitem{Frahm} K.Frahm et al., {\it Europhys.Lett.}, {\bf 49}, (2000) 48
\bibitem{Sm2} U.Smilansky in: {\it Supersymmetry and Trace Formulae:
Chaos and Disorder}, edited by I.V.Lerner et al., (Kluwer, NY, 1999),
p. 173
\bibitem{BGS} F.Borgonovi,~I. Guarneri and D.L.Shepelyansky
{\it Phys.Rev.} {\bf A 43}, 4517 (1991); F.Borgonovi and I.Guarneri
{\it Phys.Rev.} {\bf E 48} R2347 (1993)
\bibitem{Mel} P.A.Mello and J.-L.Pichard {\it J.de Phys. I} {\bf 1},
493 (1991)
\bibitem{Kol} M.Gl\"{u}ck, A.R.Kolovski and H.J.Korsch
{\it Phys.Rev.} {\bf E 60}, 247 (1999)
\bibitem{KS} K.Zyczkowski and H.-J. Sommers {\it chao-dyn/9910032}
\bibitem{abs} E. Kogan, P.A.Mello and He Liqun {\it Phys.Rev.}
{\bf E 61}( 2000) R17; C.W.J. Beenakker and P.W.Brouwer
{\it e-preprint} cond-mat/9908325
 \bibitem{Mehta} M.L. Mehta {\it Random Matrices} 2nd ed. (Academic
Press, London, 1991)
\bibitem{sec} F.Haake et al.  {\it J.Phys.A:Math.Gen.} {\bf 29}, 3641
(1996); S.Ketteman, D.Klakow and U.Smilansky {\it J.Phys.A} {\bf 30}, 3643 (1997)

\end{references}
\end{document}